\documentclass{article}
\usepackage{pdfpages}
\usepackage[english]{babel}
\usepackage{hanging}
\usepackage{amssymb}

\usepackage[letterpaper,top=2cm,bottom=2cm,left=3cm,right=3cm,marginparwidth=1.75cm]{geometry}

\usepackage{amsmath}
\usepackage{graphicx}
\usepackage[colorlinks=true, allcolors=blue]{hyperref}

\title{\Large 
{\bf Mitigating stimulated Brillouin scattering in
multimode fibers with focused output via
wavefront shaping}}
\author{%
 \parbox{\linewidth}{\centering
      Chun-Wei Chen$^{1,+}$, Linh V. Nguyen$^{2,3,4,+}$, Kabish
Wisal$^{5,+}$, Shuen Wei$^{2,+}$, \\
Stephen C. Warren-Smith$^{2,3,4,*}$, Ori Henderson-Sapir$^{2,6}$, Erik P. Schartner$^{2}$, \\ 
Peyman Ahmadi$^{7}$, Heike Ebendorff-Heidepriem$^{2}$, A. Douglas Stone$^{1,*}$, \\ David J. Ottaway$^{2,6}$, and Hui Cao$^{1,*}$%
    }%
  }%
  
\date{}

\begin{document}
	\maketitle
\begin{hangparas}{2mm}{1}
$^1$Department of Applied Physics, Yale University, New Haven, 06520, CT, USA.

$^2$Institute for Photonics and Advanced Sensing, School of Physics, Chemistry and Earth Sciences , The University of Adelaide, Adelaide, 5005, SA, Australia.

$^3$Future Industries Institute, University of South Australia, Mawson Lakes, 5095, SA, Australia.

$^4$Laser Physics and Photonics Devices Laboratoy, University of South Australia, Mawson Lakes, 5095, SA, Australia.

$^5$Department of Physics, Yale University, New Haven, 06520, CT, USA.

$^6$The Australian Research Council, Centre of Excellence for Gravitational Wave Discovery (OzGrav) 

$^7$Coherent, 1280 Blue Hills Ave., Bloomfield, 06002, CT, USA.

$^+$These authors contributed equally to this work.

$^*$Corresponding authors: Stephen.Warren-Smith@unisa.edu.au, douglas.stone@yale.edu,
hui.cao@yale.edu
 \vspace{10mm}
\end{hangparas}  
\noindent {\Large \bf Abstract} \\

\noindent The key challenge for high-power delivery through optical fibers is overcoming nonlinear optical effects. To keep a smooth output beam, most techniques for mitigating optical nonlinearities are restricted to single-mode fibers. Moving out of the single-mode paradigm, we show experimentally that wavefront-shaping of coherent input light that is incident on a highly multimode fiber can increase the power threshold for stimulated Brillouin scattering (SBS) by an order of magnitude, whilst simultaneously controlling the output beam profile. The theory reveals that the suppression of SBS is due to the relative weakness of intermodal scattering compared to intramodal scattering, and to an effective broadening of the Brillouin spectrum under multimode excitation. Our method is efficient, robust, and applicable to continuous waves and pulses. This work points toward a promising route for suppressing detrimental nonlinear effects in optical fibers, which will enable further power scaling of high-power fiber systems for applications to directed energy, remote sensing, and gravitational-wave detection. \\




\maketitle


\section{Introduction}\label{sec1}
Optical fibers facilitate nonlinear light--matter interactions through strong optical confinement and long interaction lengths~\cite{AGRAWAL2013129, stolen1974phase, digonnet1997resonantly, kobyakov2010stimulated, krupa2019multimode, wolff2021brillouin, wright2022physics}. One prominent nonlinear process is stimulated Brillouin scattering (SBS) --- light scattering mediated by acoustic phonons~\cite{AGRAWAL2013129, kobyakov2010stimulated, wolff2021brillouin}. SBS has a wide range of applications from optical phase conjugation, beam combining and cleanup, to coherent light and acoustic-wave generation, temperature and pressure sensing, and light delay and storage in fibers and integrated waveguides~\cite{rodgers1999laser, brignon2004phase, thevenaz2008slow, kim2015non, merklein2018brillouin, otterstrom2018silicon, eggleton2019brillouin, murray2022distributed}. However, SBS remains a roadblock to other applications, and many efforts have been devoted to suppressing it. SBS is a major impediment to high-power delivery of narrowband light through optical fibers and fiber amplifiers, as it converts forward-propagating light to a backward-propagating Stokes wave~\cite{pannell1993stimulated, dawson2008analysis, kobyakov2010stimulated, richardson2010high, zervas2014high, fu2017review}. Not only do the forward signals sustain significant loss, but also the backward Stokes wave can damage the upstream lasers. SBS grows from noise and limits the maximum power transmissible through the fiber.\\
    
A straightforward way of mitigating SBS is enlarging the fiber core to lower the optical intensity. However, a large core usually supports multiple guided modes, and their interference creates random speckles. Such speckled fields are often assumed to prevent generation of a clean beam suitable for applications of power delivery and coherent beam combining. Thus most techniques developed to suppress SBS are based on single-mode fibers that output a diffraction-limited beam. For example, large-mode-area microstructured fibers support single-mode operation~\cite{limpert2012yb}, but they cannot be coiled and suffer inefficient heat dissipation~\cite{limpert2003thermo, karow2012beam}. An alternative approach is tailoring the fiber geometry to modify the acoustic modes and reduce their coupling with light~\cite{kobyakov2005design, dragic2005optical, li2007ge, robin2011acoustically, Robin2014modal}. Additionally, fibers with intrinsically low nonlinearity have been fabricated~\cite{hawkins2021kilowatt}. Mass production of such specialty fibers remains a technical challenge. To use standard single-mode fibers, spectral broadening of the input light is employed to broaden the Brillouin scattering spectrum and lower the SBS growth rate~\cite{flores2014pseudo}. However, linewidth broadening is detrimental to practical applications such as coherent/spectral beam combining, dense wavelength-division multiplexing, and gravitational-wave detection~\cite{loftus2007spectrally, buikema2019narrow}. Another way of broadening the SBS spectrum is introducing large temperature or strain gradients along the fiber~\cite{yoshizawa1993stimulated, liu2007suppressing}, which is hard to implement practically and shortens the fiber lifespan. 

\begin{figure*}[h!]%
\centering
\includegraphics[width=1\textwidth]{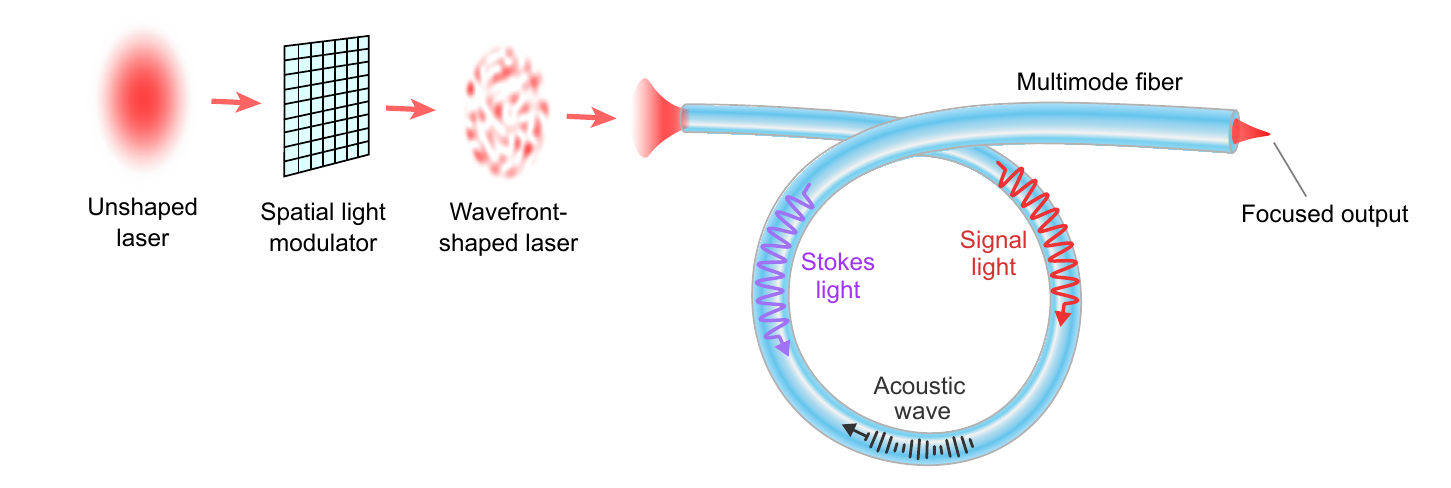}
\caption{{\bf Schematic of SBS suppression and output focusing.} Spatial wavefront of a narrowband laser beam is shaped by a spatial light modulator and excites many modes in a multimode fiber (MMF). Compared to a single-mode fiber and single-mode excitation in the same MMF, the SBS threshold power is greatly increased. Above the threshold, SBS causes a rapid increase in the power of backscattered Stokes light with the input power, while saturating the power of transmitted signal light. Input wavefront shaping modulates relative phases of fiber modes, so that their interference produces a diffraction-limited spot near the fiber output, which can be collimated by a lens. }
\label{fig:schematic}
\end{figure*}

Here we propose and demonstrate an efficient method of suppressing SBS in standard multimode fibers while maintaining narrow linewidth and high output-beam quality, via wavefront shaping.  Recent advances in wavefront-shaping techniques have enabled controlling nonlinear optical processes in multimode fibers~\cite{tzang2018adaptive, deliancourt2019wavefront, teugin2020controlling, wright2022physics, chen2022suppressing}, but suppression of SBS has not been studied previously.  Our approach is illustrated in Fig.~\ref{fig:schematic}:
By shaping the input wavefront of a narrowband signal to excite many fiber modes, we simultaneously increase the SBS threshold and focus the transmitted light to a diffraction-limited spot, which can be collimated by a lens. The SBS suppression is not simply due to reduced intensity when the core diameter is increased. The multimode excitation broadens the SBS gain spectrum and lowers the peak gain for the same intensity spread in the core. One advantage over prior schemes is that our method does not cause spectral broadening of the {\it transmitted} light, facilitating narrowband applications. We maximize the SBS threshold by optimizing the excited modal content in a multimode fiber (MMF). The selective mode excitation efficiently broadens the SBS spectrum by taking advantage of variations in Brillouin scattering strength among different fiber-mode pairs. Combining the effects of intensity reduction and coherent multimode excitation, the SBS threshold in a step-index MMF with 20-$\mu$m core is one order of magnitude higher than that of a conventional single-mode fiber ($\sim$8-$\mu$m core and $\sim$0.1 NA) of the same length \cite{AGRAWAL2013129, kobyakov2010stimulated}. Our experimental results, performed with both continuous-wave (CW) excitation at a wavelength of 1064 nm and 200-ns pulses at 1550 nm, confirm that our method is effective, robust, and generally applicable to different settings. It will enable power scaling of light delivery and amplification for various applications, e.g., directed energy and long-range remote sensing~\cite{carlson2009narrow, di2014high, Defense2019, buikema2019narrow}.



\section{Results}\label{input_offset}

\subsection{SBS suppression by multimode excitation}\label{sec:input_offset}

    \begin{figure*}[htb!]%
    \centering
    \includegraphics[width=1\textwidth]{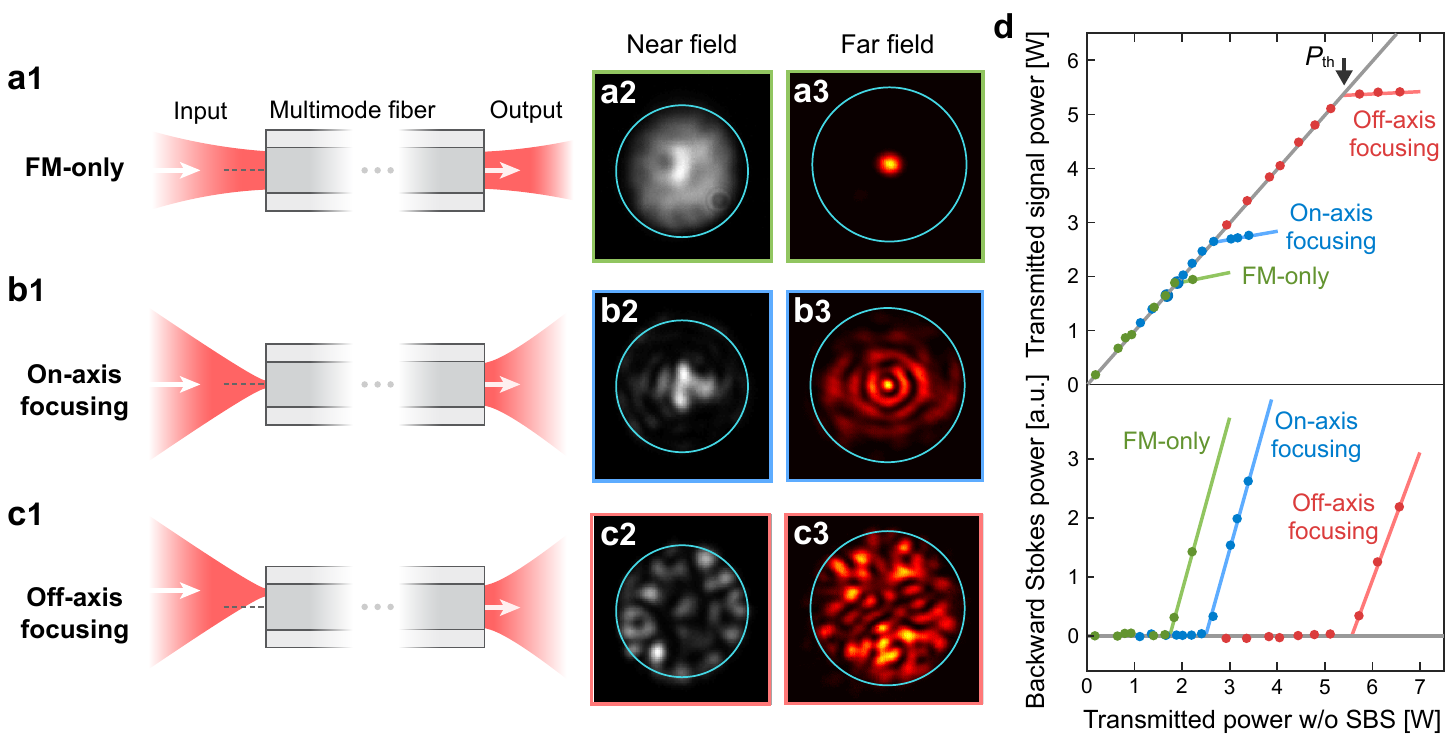}
    \caption{{\bf Single-mode vs.~multimode excitation.} {\bf a1}, Input signal is injected only to the fundamental mode (FM), producing a single spot in the far-field intensity pattern at multimode fiber output ({\bf a3}). {\bf b1}, Tight focusing of input light to fiber axis excites both fundamental and high-order radial modes, outputting a multiple-ring pattern in the far field ({\bf b3}). {\bf c1}, Off-axis focusing excites additional high-order non-radial modes, creating a speckled far-field pattern ({\bf c3}). {\bf a2--c2}, Measured near-field intensity patterns at fiber output reveal comparable transverse spreading across the core for three excitation schemes. {\bf d}, Measured transmitted signal power (upper) and backscattered Stokes power (lower) vs.~expected transmitted power without SBS. Above SBS threshold, transmitted signal power saturates, while reflected Stokes power increases rapidly with input power. SBS threshold power increases from 1.8 W with FM-only excitation, to 2.6 W for on-axis focusing, and finally 5.4 W for off-axis focusing. }

    \label{fig:3exct}
    \end{figure*}

    We first demonstrate SBS mitigation by coherent multimode excitation. To this end, we vary the number of excited modes and measure the SBS threshold. The input signal (serving as the pump for SBS) is a linearly-polarized continuous wave from a narrowband (15-kHz) fiber laser at 1064 nm. It is launched into a 50-meter-long, Ge-doped, step-index MMF with $\sim$20-$\mu$m core diameter and 0.3 numerical aperture (NA), supporting $\sim$80 modes per polarization. The transmitted light from the distal fiber end is split to simultaneously measure the output power and near-/far-field intensity patterns.  We also monitor the input and backscattered light at the proximal fiber end (Supplementary Information, sec.~1). \\ 

    We start with exciting only the fundamental mode (FM) in the MMF by focusing the incident light to the proximal fiber facet using a lens of NA matching that of the FM [Fig.~\ref{fig:3exct}a]. The transmitted beam produces a single spot at the center of the far field. We gradually increase the input power, and the output power first increases linearly but then levels off [Fig.~\ref{fig:3exct}d]. Meanwhile, the reflected power grows rapidly, indicating the onset of backward SBS. The transmitted power at this reflection threshold is 1.8 W. \\
   	
	Next, we excite multiple modes in the MMF by switching to a lens of larger NA (close to the core NA) [Fig.~\ref{fig:3exct}b]. The input beam is focused onto a small spot at the proximal fiber facet. When the focal spot is at the core center, predominantly the FM and several purely radial modes are excited. The output far-field pattern displays concentric rings. Due to modal interference, the near-field pattern is smaller than that of FM-only excitation. Nevertheless, the SBS threshold power rises to 2.6 W, indicating that the SBS suppression is not caused by transverse energy spreading that lowers the intensity. Instead, it is due to spectral broadening and peak reduction of the SBS gain by multimode excitation (see sec.~\ref{sec:theory}). \\
	
	To excite even more modes, we move the focal spot away from the core center [Fig.~\ref{fig:3exct}c]. Additional higher-order modes (HOM) with nonzero azimuthal index are excited in addition to the FM and radial HOMs. Consequently, the output beam is highly speckled. The SBS threshold increases with the distance $d_{\rm in}$ of the focal spot from the fiber core center [Fig.~\ref{fig:theory}a], as more modes are excited. Near the core edge, the threshold reaches 5.4 W, $\sim$3 times the FM-only threshold.

\subsection{Broadening of Brillouin scattering spectrum} \label{sec:theory} 
   
    To understand the mechanism of SBS suppression with multimode excitation, we have developed a multimode SBS theory (detailed derivation in \cite{wisal2023SBStheory}). Consider Brillouin scattering of a forward-propagating photon (signal) in fiber mode $l$ to a backward photon (Stokes) in mode $m$ via emission of a forward acoustic phonon at frequency $\Omega$. The Stokes photon has a frequency shift of $\Omega$ from the signal photon. The scattering strength $g_{\rm B}^{(m,l)}(\Omega)$ is determined by the spatial overlap of optical modes $m$ and $l$ with acoustic modes in the fiber~\cite{wisal2023SBStheory}. We order the modes by their propagation constants from high to low. \\

    \begin{figure*}[htb!]
    \centering
    \includegraphics[width=1\textwidth]{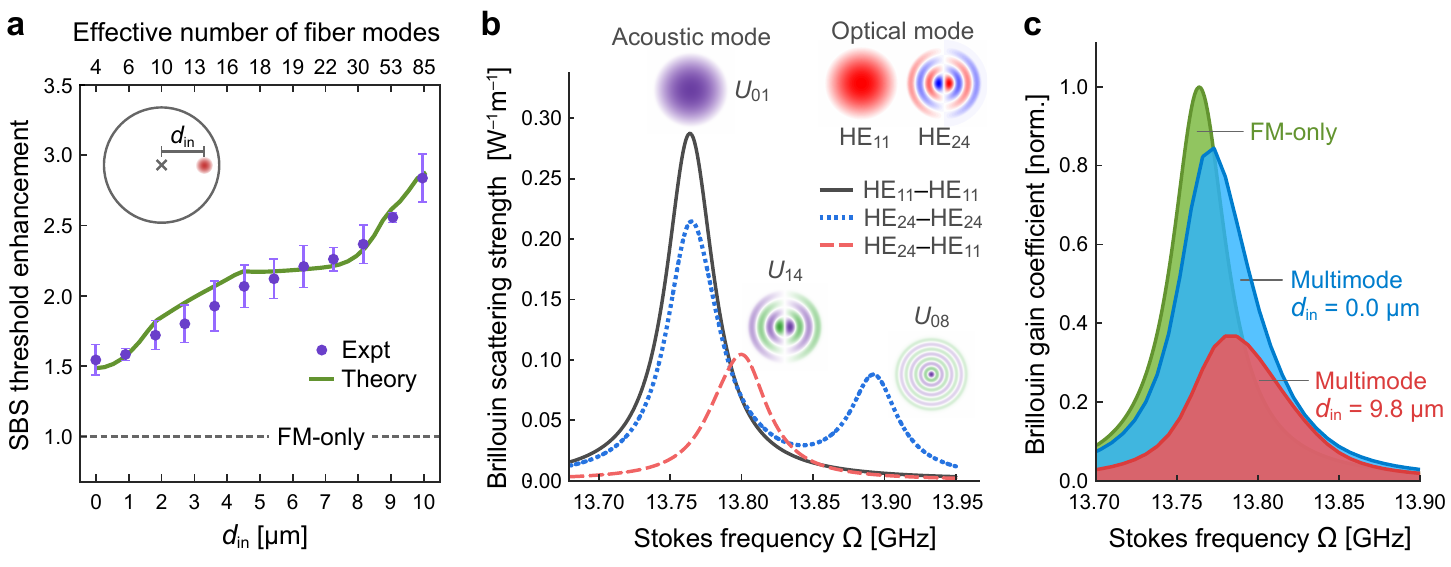}
    \caption{{\bf Brillouin gain spectrum and SBS threshold.} {\bf a,} Experimentally measured and theoretically predicted SBS threshold enhancement over FM-only excitation increases with distance $d_{\rm in}$ of the focused input beam to fiber axis (inset; spot size $\approx$ 3.4 $\mu$m). Dots: mean enhancement of experimental data, error bars: standard deviation, solid curve: theoretical prediction including mode-dependent loss, mode coupling, and polarization mixing.
    {\bf b,} Calculated Brillouin scattering strength $g_{\rm B}^{(m,l)}(\Omega)$ for two optical modes HE$_{\rm 11}$ and HE$_{\rm 24}$ (each mode profile displays the $x$-polarized field component).  Intramodal and intermodal Brillouin scattering have peaks at different Stokes frequency $\Omega$, and the corresponding acoustic modes are displayed above the peaks.  
    {\bf c,} Calculated Brillouin gain spectrum $G_{\rm B}^{(m)}(\Omega)$ for FM-only excitation (green) is narrower than multimode excitation with on-axis focusing of input light (blue).  Off-axis focusing increases the number of excited modes and further broadens the Brillouin gain spectrum (red); the resulting decrease in peak gain enhances the SBS threshold. 
    }
    \label{fig:theory}
    \end{figure*}
    
    We refer $m = l$ to intramodal scattering and $m \neq l$ to intermodal scattering. Figure~\ref{fig:theory}b shows that if the signal is in the FM ($l$ = 1), the SBS is the strongest for Stokes in the FM ($m$ = 1), with scattering strength equal to
    $g_{\rm B}^{(1,1)}(\Omega)$, where the peak Stokes frequency $\Omega$ equals the eigenfrequency of the lowest-order acoustic mode. The intramodal scattering for an HOM (e.g., HE$_{\rm 24}$) has a lower peak at the same $\Omega$, due to smaller overlap with the lowest-order acoustic mode, and an additional peak at higher $\Omega$ corresponding to a higher-order acoustic mode. Intermodal scattering is generically weaker than intramodal scattering, due to smaller acousto-optic overlap for non-identical signal and Stokes mode profiles. The intermodal peak is not only lower, but also spectrally shifted from the intramodal peak, because the acoustic mode having the largest spatial overlap with the signal and Stokes modes is a higher-order mode with higher eigenfrequency. \\
    
    Seeded by spontaneous Brillouin scattering from mode $l$ to $m$, the Stokes power $P^{\rm s}_m$ grows exponentially via SBS while propagating backward in the fiber. The growth rate is given by the product of the scattering strength $g_{\rm B}^{(m,l)}(\Omega)$ and the signal power $P_l$~\cite{wisal2023SBStheory}. The SBS threshold is defined by the transmitted signal power at which a significant fraction of the input power is lost to backward Stokes. Below the SBS threshold, however, the Stokes power is at most a few percent of the input power, thus the depletion of signal power by SBS is negligible. If the signal power is distributed among $M$ modes, the growth rate for $P^{\rm s}_m$ is given by the weighted sum $G_{\rm B}^{(m)}(\Omega) = \sum_{l=1}^{M}{g_{\rm B}^{(m,l)}(\Omega) {P}_l}$, defined as the Brillouin gain coefficient. The Stokes power, which is maximal at the fiber proximal end ($z=0$), is:
    \begin{equation}
    P^{\rm s}_m(\Omega, z=0)=P^{\rm s}_m(\Omega, z=L) e^{G_{\rm B}^{(m)}(\Omega) L} 
    ,
    \label{eq1}
    \end{equation} 
\noindent where $P^{\rm s}_m(\Omega, z=L)$ is the seed from spontaneous Brillouin scattering at the fiber distal end~\cite{kobyakov2010stimulated}. The exponential growth rate of Stokes power is dictated by the peak value of $G_{\rm B}^{(m)}(\Omega)$. \\
    
Let us analyze $G_{\rm B}^{(m)}(\Omega)$ for a few cases. When only the FM is excited by the signal in a MMF, the Brillouin gain is strongest for the Stokes in the FM [$G_{\rm B}^{(1)}(\Omega)=g_{\rm B}^{(1,1)}(\Omega) P_1$]. Since $g_{\rm B}^{(1,1)}(\Omega)$ has the highest peak among all $g_{\rm B}^{(m,l)}(\Omega)$ [Fig.~\ref{fig:theory}b], it results in the lowest SBS threshold.
Instead, if a single HOM ($l \neq 1$) is excited, Stokes growth in that HOM ($m =l$) is strongest. Since intramodal scattering for an HOM is weaker than that for the FM, the SBS threshold is slightly higher. A more dramatic increase of the SBS threshold occurs when multiple modes are excited by the signal. Since the Brillouin gain $G_{\rm B}^{(m)}(\Omega)$ is a power-weighted sum of intramodal and intermodal scattering, it is spectrally broadened for Stokes in individual mode $m$ due to frequency-offset peaks of $g_{\rm B}^{(m,l)}$ among different mode pairs ($m$,$l$). Moreover, intermodal peaks are lower than intramodal ones [Fig.~\ref{fig:theory}b]. 
Consequently, the peak growth rate of Stokes power is greatly reduced, leading to significant threshold enhancement over single-mode excitation. \\  
        
To compare with experimental data [Fig.~\ref{fig:theory}a], we calculate the modes excited by a signal focused on the proximal facet and obtain the Brillouin gain spectrum and the SBS threshold (Supplementary Information, sec.~2.1). A tight focus on the core center excites multiple modes, leading to a broader Brillouin gain spectrum than FM-only excitation and thus a lower peak [Fig.~\ref{fig:theory}c]. Shifting the focus transversely to excite even more modes, the spectrum is further broadened with a much lower peak. The SBS threshold, predicted by our theory, increases monotonically with $d_{\rm in}$, and agrees well with the experimental data [Fig.~\ref{fig:theory}a]. To obtain such agreement, we experimentally characterize the mode-dependent loss, linear mode coupling, and polarization mixing of our MMF, and include these effects in the SBS theory (Supplementary Information, sec.~2.2). Slope variation in the data is reproduced and explained by the theory (Supplementary Information, sec.~2.1). \\

\subsection{Robustness of suppression scheme}\label{random}

    \begin{figure*}[b!]%
    \centering
    \includegraphics[width=1\textwidth]{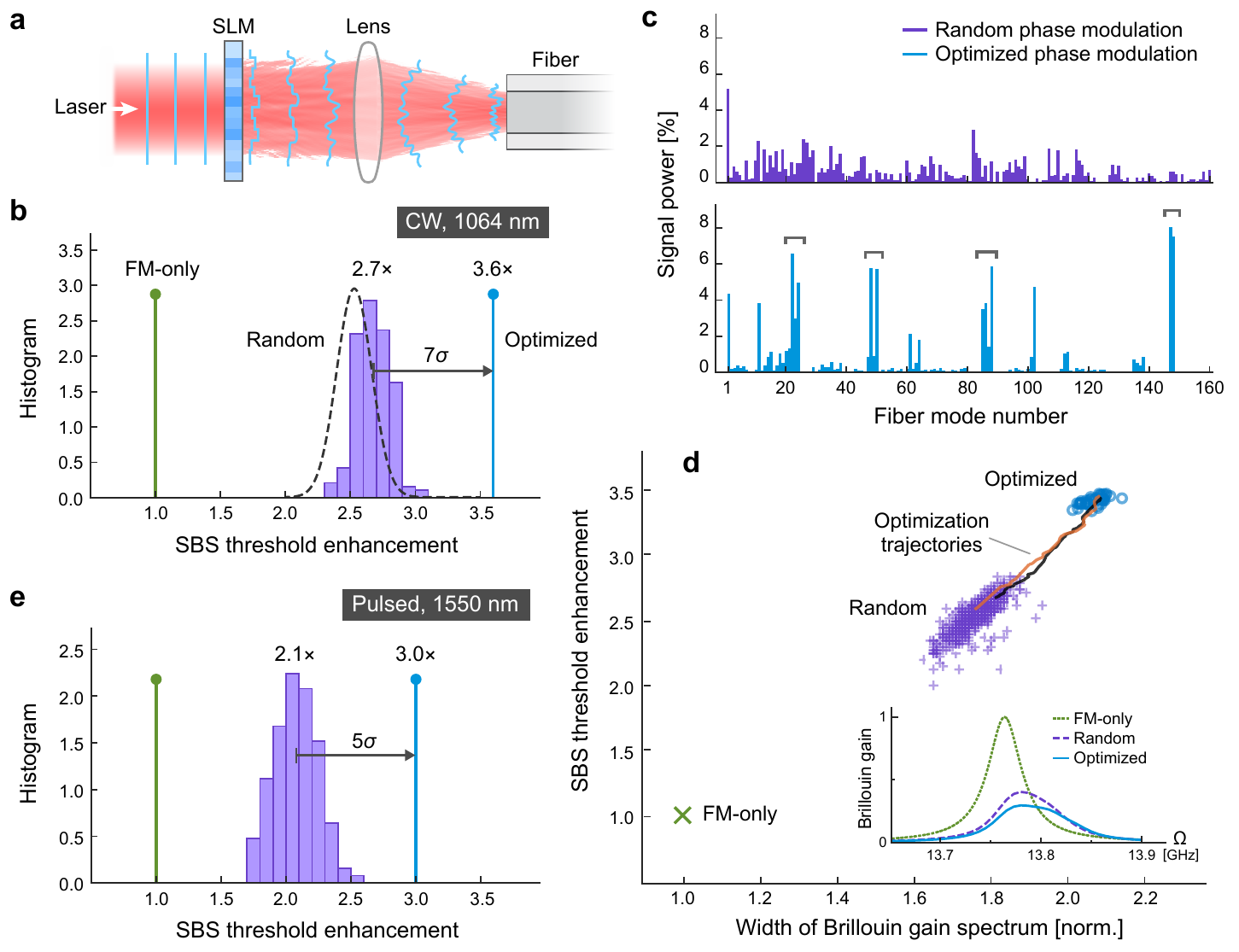}
    \caption{{\bf Robustness and optimization of SBS suppression by wavefront shaping.} 
    {\bf a,} Wavefront shaping of a laser beam by phase-only SLM at conjugate plane of proximal fiber facet. 
    {\bf b,} Histogram of SBS threshold enhancement for 190 random phase patterns (purple) with CW laser at 1064 nm. Mean enhancement $\sim$2.7 over FM-only excitation (green) and standard deviation $\sigma \approx 0.14$ agree well with theoretical prediction (dashed).  Optimization of phase pattern further increases SBS threshold enhancement to 3.6 (blue).    
    {\bf c,} Simulated signal mode content is relatively uniform with random phase modulation (upper), but concentrated in several widely-spaced mode groups with optimized phase modulation (lower). 
    {\bf d,} SBS threshold increases with spectral width of Brillouin gain. Inset: Brillouin gain spectra for FM-only excitation, multimode excitation by random and optimized phase modulations (FWHMs: 40, 72, 85 MHz). 
    {\bf e,} Similar to {\bf b} with 200-ns pulses at 1550 nm. Mean enhancement by random (optimized) phase modulation is $\sim$2.1 (3.0).    
    }
    \label{fig:robust}
    \end{figure*}

The experiment with input focusing validates our multimode scheme for SBS suppression, but is restricted to considering a focused Gaussian input. To examine how robust this method is, we conduct a statistical study of the SBS threshold for arbitrary input wavefronts. We imprint random phase patterns on the signal beam (CW at 1064 nm) with a spatial light modulator (SLM) [Fig.~\ref{fig:robust}a]. Strikingly, for all random input wavefronts, the threshold enhancement over FM-only excitation is consistently $> 2$ [Fig.~\ref{fig:robust}b]. The histogram shows a mean enhancement of $\sim$2.7 and a small standard deviation $\sigma \approx 0.14$. The prediction from simulations based on our theory is consistent with the experimental data [Fig.~\ref{fig:robust}b]. It confirms that the random input wavefront distributes the signal power rather uniformly among all fiber modes [upper panel, Fig.~\ref{fig:robust}c]. The calculated Brillouin gain spectrum [inset, Fig.~\ref{fig:robust}d] has a full-width-at-half-maximum (FWHM) of 72 MHz, wider than the FM-only FWHM of 40 MHz. Consequently, the gain peak is notably lower, leading to threshold increase. \\

Figure~\ref{fig:robust}d reveals the correlation between the threshold enhancement and Brillouin gain spectrum width. Different random input wavefronts cause variations in the excited mode content, leading to slightly different widths of the Brillouin gain spectrum. Larger widths correspond to higher SBS thresholds, with a Pearson correlation coefficient of 0.78. \\
    
To demonstrate the generality of our approach for SBS suppression, we switch from CW to pulsed (200-ns) signal at a different wavelength (1550 nm) and repeat the above experiments on the same type of MMF [Fig.~\ref{fig:robust}e]. Under random multimode excitation, the mean SBS threshold enhancement is 2.1, slightly lower than that at 1064 nm. This is expected at a longer wavelength, where the fiber supports fewer optical modes ($\sim$37 per polarization). \\

\subsection{Optimization of multimode excitation}

The input-focusing data and the spread of SBS thresholds under random input wavefronts suggest that some multimode combinations are more efficient than others in mitigating SBS. These results prompt us to optimize the input wavefront for further enhancement of the SBS threshold. We sequentially optimize the phases of 8$\times$8 SLM macropixels, to increase the transmitted signal power and/or decrease the backscattered Stokes power (Supplementary Information, sec.~1.4). The threshold is increased to 3.6 times the FM-only threshold, higher than that with off-axis input-focusing, and $\sim$$7\sigma$ higher than the mean enhancement with random wavefronts [Fig.~\ref{fig:robust}b]. To better understand the mechanism underlying the threshold enhancement, we combine our theory with a numerical simulation of the SLM phase optimization to maximize the SBS threshold. The threshold enhancement reaches 3.5, closely matching the experimental result (Supplementary Information, sec.~2.3). \\
    
As shown in Fig.~\ref{fig:robust}c, the optimized mode content (from simulation) is not uniform; instead the power is concentrated in several groups that are widely spaced in propagation constant. Such selective mode excitation broadens the Brillouin gain spectrum more effectively than uniform mode excitation, and the FWHM reaches 85 MHz [inset, Fig.~\ref{fig:robust}d]. Starting from different input wavefronts, the optimization procedure takes different routes, and the final SBS thresholds vary slightly [Fig.~\ref{fig:robust}d]. Although the optimized mode contents differ from one another, they all consist of well-separated groups for maximal broadening of the Brillouin gain spectrum.  \\

\subsection{Control of output beam profile} \label{output_focusing}

Finally, we show that wavefront shaping of a coherent signal not only can mitigate the SBS in an MMF, but also can shape the transmitted beam profile. Since our theory finds that the SBS suppression depends only on how the signal \emph{power} is distributed among modes [Eq.~\ref{eq1}],  the $phase$ of the input field to individual modes can be adjusted to control the modal interference at the fiber output. Below we present an example of focusing the output light to a diffraction-limited spot. Perfect focusing requires full control of the input field amplitude and phase~\cite{gomes2022near}, whereas we have phase-only control with an SLM.  Nevertheless, we find that this incomplete control can still direct most transmitted power to a designated location and simultaneously increase the SBS threshold.  \\
    
    \begin{figure*}[hbt!]%
    \centering
    \includegraphics[width=1\textwidth]{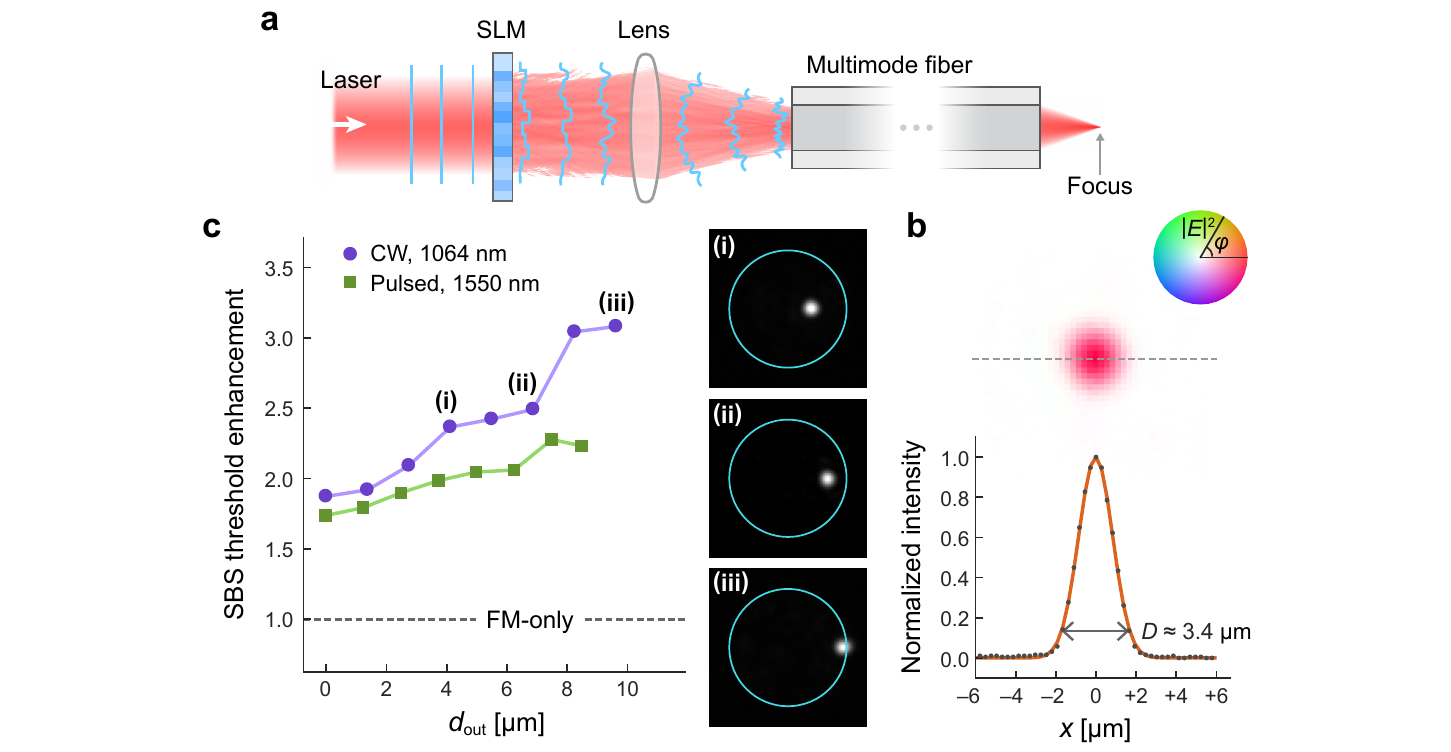}
    \caption{{\bf SBS suppression with focused output.} 
    {\bf a,} Simplified setup of wavefront shaping to focus the transmitted light to a diffraction-limited spot near the fiber distal facet. {\bf b,} Measured focal spot with flat phase (phase expressed by hue and intensity by saturation in the color wheel). Its $e^{-2}$ diameter $D \approx 3.4$ $\mu$m is close to the diffraction limit ($\approx 3.3$ $\mu$m). {\bf c,} SBS threshold enhancement (over FM-only excitation) increases monotonically with distance $d_{\rm out}$ between the output focal spot and fiber axis, for both CW at 1064 nm (purple circles) and 200-ns pulses at 1550 nm (green squares). Inset: output intensity profiles for focusing CW at $d_{\rm out} =$ 4.1 $\mu$m (i), 6.9 $\mu$m (ii), 9.5 $\mu$m (iii). }
    \label{fig:outfocus}
    \end{figure*}
    
As illustrated in Fig.~\ref{fig:outfocus}a, a linearly-polarized Gaussian beam is phase-modulated by an SLM at the conjugate plane of the fiber input facet, so that the transmitted beam is focused to a spot near the output facet. We optimize the phases of 16$\times$16 macropixels sequentially to maximize the output intensity at the focus~\cite{vellekoop2008phase} (Supplementary Information, sec.~1.4). The focal spot can be placed anywhere within the field of view defined by the fiber core size and NA. We choose the focus to be 20 $\mu$m outside the output facet (in the air) to avoid damage to the fiber. Figure~\ref{fig:outfocus}b shows that, upon optimization, the focal spot size ($D \approx 3.4$ $\mu$m at $e^{-2}$ of the maximum intensity) is almost at the diffraction limit ($D \approx 3.3$ $\mu$m) that is determined by the fiber core NA $\approx 0.3$. We also measure the phase of the transmitted field, and the flat phase across the focal spot confirms the diffraction-limited focusing. The power within the focus is $\sim$0.7 of the total output power, close to the limit for phase-only modulation of a Gaussian beam (Supplementary Information, sec.~2.4).  \\     

Since the formation of a tight focus comes from the interference of many fiber modes, multimode excitation leads to SBS suppression. We measure the SBS thresholds when focusing the output light at different distances $d_{\rm out}$ from the fiber axis. On-axis focusing ($d_{\rm out} = 0$) results in a 1.8$\times$ increase of the threshold over FM-only excitation [Fig.~\ref{fig:outfocus}c]. Moving the focus away from the fiber axis further increases the threshold, up to 3.1$\times$ enhancement at $d_{\rm out} \approx 9.5$ $\mu$m. This is because forming an off-axis focus needs an increased number of participating modes, e.g., $> 70$ modes for focusing at $d_{\rm out} \approx 9.5$ $\mu$m. We also demonstrate this with 200-ns laser pulses at 1550 nm, confirming the broad applicability of our method for simultaneously achieving a high SBS threshold and controlling the output-beam profile.


\section{Discussion}\label{discussion}
    
Wavefront shaping enables high-power delivery through an MMF with a smooth output beam. While the SBS threshold with multimode excitation exceeds 3$\times$ of the FM-only excitation in the same MMF, we note that the FM-only threshold in our fiber is already $\sim$4$\times$ higher than that in a typical single-mode fiber. The SBS threshold and focusing quality can be further improved by gaining full control over all the fiber modes, requiring amplitude and phase modulation of input light for both polarizations. Our theory predicts the SBS threshold enhancement of 4.5$\times$ in our MMF with complete control of a narrowband input signal (Supplementary Information, sec.~2.3). With output beam control, our method allows utilizing highly multimode fibers with even larger cores (diameter $> 100$ $\mu$m), leading to orders of magnitude increase in threshold \cite{wisal2023SBStheory}. In terms of output focusing, near-unity ($> 95\%$) power concentration in the focal spot has been achieved with complete control of input fields to an MMF~\cite{gomes2022near}. Since direct amplitude modulation introduces power loss, lossless full-field shaping can be realized using two phase-only SLMs with a distance between them~\cite{jesacher2008near}. The SLM patterns need to be constantly adjusted to stabilize the beam profile against temporal drift and fluctuations of the fiber. In our current experiment, the 50-meter-long MMF, loosely coiled and placed on an optical table without temperature stabilization and mechanical isolation, drifts on the time scale of 0.5--3 hours. The wavefront optimization currently takes $\sim$12 minutes, while recent works show output refocusing in less than a second~\cite{caravaca2013real, feldkhun2019focusing}. It is also noteworthy that an output beam profile other than a focal spot can be obtained by tailoring the input wavefront~\cite{vcivzmar2011shaping,leite2018three} \\

While we demonstrate the SBS suppression in passive MMFs, our scheme is applicable to MMFs with optical gain, providing a robust route toward further power scaling for high-power fiber amplifiers. Furthermore, focusing an amplified output to a diffraction-limited spot by input wavefront shaping has been realized experimentally~\cite{florentin2017shaping}. In addition, our approach can suppress other nonlinear effects such as stimulated Raman scattering~\cite{tzang2018adaptive} and transverse mode instability~\cite{chen2022suppressing}. \\
    
Finally, the proof-of-concept experiments are conducted on standard MMFs, but our method is readily adopted for specialty fibers and combined with other schemes of SBS suppression~\cite{kobyakov2005design, dragic2005optical, li2007ge, robin2011acoustically, Robin2014modal, hawkins2021kilowatt}. We envision that wavefront shaping can be a powerful tool to suppress multiple detrimental nonlinear effects while outputting a desired beam profile. \\

\section{Acknowledgments}

We thank Yaniv Eliezer at Yale University and Hasan Y{\i}lmaz at Bilkent University for stimulating discussions, Nicholas Bernardo and Vincent Bernardo at Yale University for technical assistance. We acknowledge the computational resources provided by Yale University and the University of Adelaide. This work was performed in part at the Opto Fab node of the Australian National Fabrication Facility supported by the Commonwealth and SA State Government. This work is supported by the Air Force Office of Scientific Research (AFOSR) under Grants FA9550-20-1-0129 and FA9550-20-1-0160, the Australian Research Council under CE170100004, and the Next Generation Technologies Fund (NGTF) [Research Agreement 10737] through the NGTF Directed Energy (DE) Science and Technology (S\&T) Network of Australian Universities and Industry Partners. The Australian authors acknowledge the support received from the Commonwealth of Australia. Stephen C. Warren-Smith is supported by an Australian Research Council (ARC) Future Fellowship (FT200100154). \\

\section{Author contributions}

H.C. proposed the idea and initiated this project; C.-W.C. performed the continuous-wave experiments using the fiber amplifier built by P.A. and analyzed the data in collaboration with K.W., under the supervision of H.C.; L.V.N. and S.W. performed the pulsed experiments and analyzed the data in collaboration with S.C.W.-S., O.H.-S., and E.P.S., under the supervision of H.E.-H and D.J.O.; K.W. developed the theory and numerical simulations in collaboration with S.C.W.-S. and C.-W.C., under the supervision of A.D.S.; C.-W.C., L.V.N., K.W., A.D.S., and H.C. wrote the manuscript with input from all authors. \\

\bibliographystyle{unsrt}
\bibliography{sn-bibliography}



\section*{Supplementary material}

# Supplementary Information
# Mitigating stimulated Brillouin scattering in multimode fibers with focused output via wavefront shaping

Chun-Wei Chen[1,+], Linh V. Nguyen[2,3,4,+], Kabish Wisal[5,+], Shuen Wei[2,+],
Stephen C. Warren-Smith[2,3,4,*], Ori Henderson-Sapir[2,6], Erik P. Schartner[2],
Peyman Ahmadi[7], Heike Ebendorff-Heidepriem[2], A. Douglas Stone[1,*],
David J. Ottaway[2,6], and Hui Cao[1,*]

[1]Department of Applied Physics, Yale University, New Haven, 06520, CT, USA.
[2]Institute for Photonics and Advanced Sensing, School of Physics, Chemistry and Earth Sciences, The University of Adelaide, Adelaide, 5005, SA, Australia.
[3]Future Industries Institute, University of South Australia, Mawson Lakes, 5095, SA, Australia.
[4]Laser Physics and Photonics Devices Laboratoy, University of South Australia, Mawson Lakes, 5095, SA, Australia.
[5]Department of Physics, Yale University, New Haven, 06520, CT, USA.
[6]The Australian Research Council, Centre of Excellence for Gravitational Wave Discovery (OzGrav)
[7]Coherent, 1280 Blue Hills Ave., Bloomfield, 06002, CT, USA.
[+]These authors contributed equally to this work.
[*]Corresponding authors: Stephen.Warren-Smith@unisa.edu.au, douglas.stone@yale.edu, hui.cao@yale.edu

## 1 Experiment

### 1.1 Optical setup

We verify our scheme of SBS suppression by multimode excitation in two separate experiments. One is conducted with continuous waves (CW) at $\lambda = 1064$ nm, the other with 200-ns pulses at $\lambda = 1550$ nm.

Figure S1a depicts the first experimental setup. A fiber amplifier (research unit from Coherent Nufern) seeded by a CW fiber laser with a linewidth of 15 kHz (NP Photonics Rock 1$\mu$m) produces the linearly polarized signal (pump for Brillouin scattering). It is collimated by a lens and passes through an optical isolator to prevent the laser amplifier from being damaged by strong backscattered light from SBS. We use a zero-order half-wave plate (HWP) and a Glan-Taylor polarizer to control the signal power, a $4f$ system for beam expansion, and a phase-only spatial light modulator (SLM) for wavefront shaping. The signal beam is expanded to cover the active area of the SLM (Meadowlark HSP1920-500-1200-HSP8 with a water cooling system) to prevent damage at high power. The reflected light propagates through another $4f$ system with a pinhole at the focal plane to filter the zeroth-order diffraction from the SLM. It is then split by a non-polarizing beam splitter (BS), and the reflected light (10% of total power) is directed to a photodetector (PD) to monitor the input power. The transmitted light (90% of total power) is launched into a multimode fiber (MMF) by an objective lens (OL, $\mu$-Spot LMH-20X-1064) for multimode excitation or a plano-convex lens (Thorlabs LA4725-1064) with a longer focal length ($f = 75$ mm) for fundamental-mode-only excitation.

The MMF (Coherent Nufern FUD-3607) is 50 meters long and loosely coiled on an optical table without active cooling and mechanical isolation. The fiber core is germanium-doped, 20 $\mu$m in diameter, and has a numerical aperture (NA) of $\sim$0.3. The number of guided modes at the signal wavelength $\lambda = 1064$ nm is 80 per polarization. The transmitted light from the fiber is collected by an objective identical to that at the fiber input, and divided by beam splitters to several beams for

simultaneous measurement of near-field and far-field intensity patterns, total power, and its temporal fluctuation. The near- and far-field patterns are captured by near-infrared cameras (Allied Vision Mako GM-419B-NIR and Xenics Xeva-1.7-320 TE3). The time-integrated power is measured by a slow PD (Newport 818-SL), while the temporal fluctuation on the sub-$\mu$s scale is recorded by a fast PD (Thorlabs DET10A or PDA20CS). Similarly, at the fiber input end, we measure the far-field intensity pattern of backscattered light, its time-integrated power and temporal fluctuation.

In addition to the output intensity profile, we also measure the phase pattern of the near field at the fiber distal end [Fig. 5b in the main text] in an off-axis holographic setup shown schematically in Fig. S1d. A fraction of the CW seed laser with a flat phasefront serves as a reference beam, and interferes with the transmitted light from the MMF. The two beams are incident onto a camera at an angle, and their interference pattern is recorded by the camera. From it, we extract the phase pattern of the transmitted field. Figure 5b of the main text shows the measured phase across the focal spot near the fiber output end is flat, confirming diffraction-limited focusing through the MMF by shaping the input wavefront.

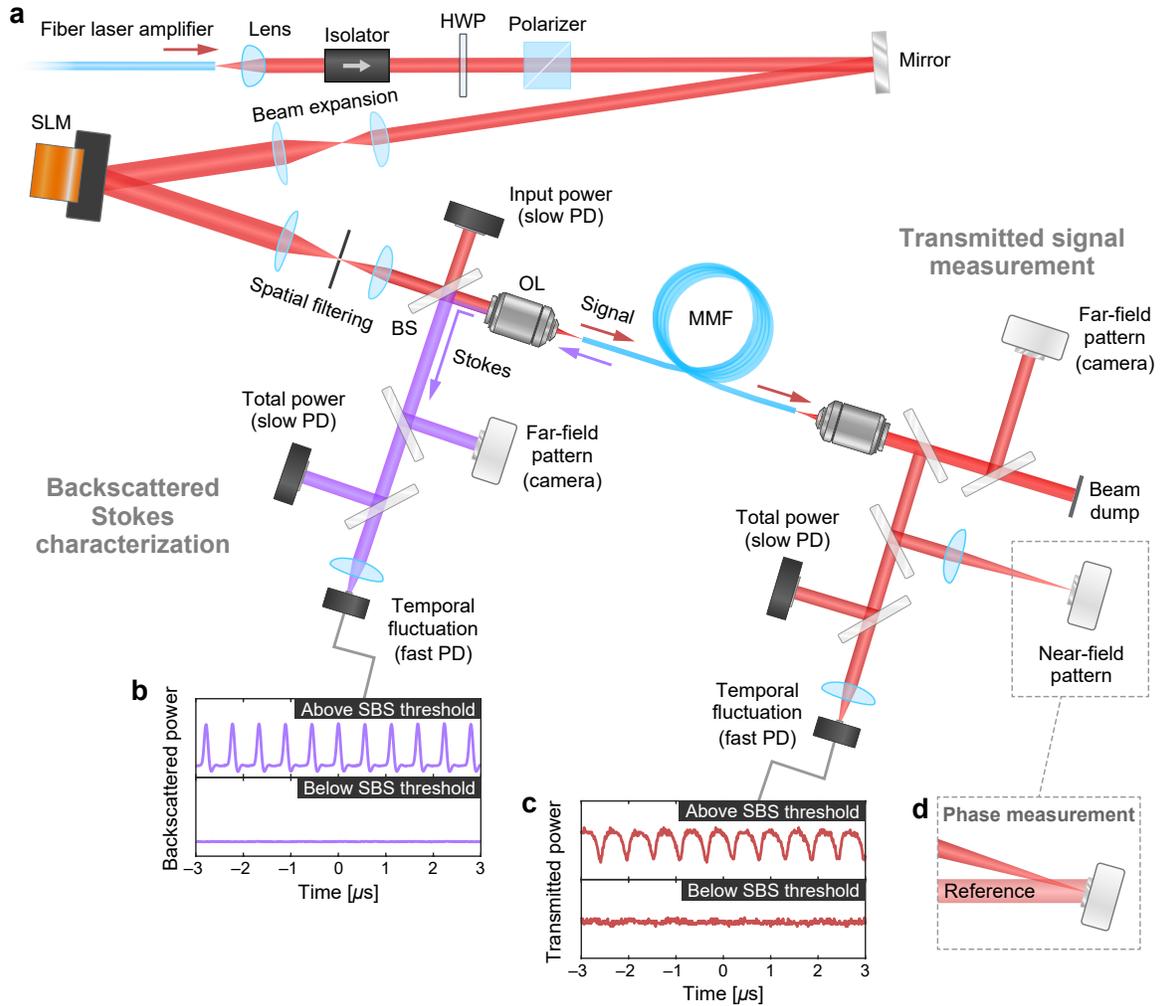

Fig. S1. Experimental setup for investigating SBS in a multimode fiber by wavefront shaping. **a,** Schematic showing the setup for wavefront shaping, input excitation, and output characterization. HWP: half-wave plate, SLM: phase-only spatial light modulator, BS: non-polarizing beam splitter with 10 % reflectivity, OL: objective lens with NA ≈ 0.3, PD: photodetector. The slow PD has a response time of ∼ms, and the fast PD of ∼10 ns. **b,c,** Measured time traces of backscattered (**b**) and transmitted (**c**) powers from a 50-meter-long MMF under CW excitation, above (upper) and below (lower) SBS threshold. SBS manifests as micro-second-scale pulsations of backward Stokes and forward signal. **d**, Off-axis holographic setup for phase measurement.

2

In the second experiment, we use a pulsed fiber laser (KEOPSYS, PEFL-E07-LP-040-200-010-W00-G3-T1-ET1-PE30D-CIRFA) that produces 200-ns pulses at a repetition rate of 10 kHz. The optical setup is similar to that of the first experiment with a CW laser. The SLM is Santec SLM-200, and an aspheric lens (Thorlabs C220TMD-C) is employed to couple light into the fiber. A quarter-wave plate (QWP) is inserted in the optical path between the SLM and the objective lens to convert the polarization state of light from linear to circular. A 50-meter-long MMF of the same core diameter and NA is used. It supports 37 modes per polarization at $\lambda = 1550$ nm. The time-integrated power of the transmitted pulses is measured by a power meter (Thorlabs S146C) with response time $< 1$ $\mu$s.

## 1.2 Characterization of SBS

The SBS threshold is determined from the dependence of the transmitted power on the input power. As shown in the upper panel of Fig. 2d in the main text, the time-integrated transmitted power first increases linearly with the input power, and then saturates at a certain power, which is set as the SBS threshold $P_{\text{th}}$. The horizontal axis of Fig. 2d is the transmitted power in the absence of SBS, which is obtained from the input power and the MMF transmittance for a given modal content (determined by the launching condition of incident light). The SBS threshold is confirmed by the power variation of backscattered light. Below the threshold, the backscattered power, due to Fresnel reflection from the proximal fiber facet and Rayleigh scattering in the fiber, rises gradually with the input power. Once the transmitted power starts to level off, the reflected power increases rapidly. In the lower panel of Fig. 2d, we subtract the power of linear backscattering (extrapolated from linear fit of the data below the SBS threshold) from the total backscattered power to show the Stokes power.

At the onset of SBS, both transmitted signal and reflected Stokes exhibit temporal fluctuations under CW excitation. Figure S1b,c shows periodic spikes of Stokes power and synchronized dips in transmitted signal power above the SBS threshold. The modulation period is about 0.5 $\mu$s, corresponding to the round-trip time of light propagation in the fiber [1, 2].

## 1.3 Single-mode vs. multimode excitation

Excitation of only the fundamental mode (FM) is achieved by using a plano-convex lens of NA $\approx 0.06$ to focus the input light to the fiber proximal facet at normal incidence. The focal spot size is roughly equal to the fundamental mode field diameter of the MMF. Under such excitation, the transmitted light from the MMF shows a single spot at the center of the far field, as expected for the FM [Fig. 2a3 in the main text]. The corresponding near-field intensity distribution is relatively smooth across the fiber core with low-contrast modulations from a small amount of higher-order modes (HOM), due to imperfect FM excitation at the input and/or linear mode coupling in the fiber [Fig. 2a2 in the main text]. Above the SBS threshold $P_{\text{FM}} \approx 1.8$ W, the backward Stokes beam profile is identical to that of the transmitted signal, indicating that SBS occurs predominantly in the FM.

To excite multiple modes in the fiber, we replace the lens with an objective lens of NA $\approx 0.3$ (close to the NA of the MMF core). The focal spot on the proximal fiber facet is $\sim 3.4$ $\mu$m, much smaller than the input beam size for FM-only excitation. In Figs. 2 and 3a of the main text, we shift the focus transversely across the fiber facet by displaying a linear phase ramp on the SLM. We gradually vary the ramping magnitude and orientation to control the deflection angle and direction, respectively. The SBS threshold increases with the distance $d_{\text{in}}$ of the focal spot from the fiber axis.

## 1.4 SLM phase modulation

To obtain the data in Fig. 4 of the main text, we use the SLM to imprint a random phase pattern on the input beam to the MMF. The SLM is placed at the conjugate plane of the proximal fiber facet. The active area of the SLM (Meadowlark HSP1920-500-1200-HSP8) is $10.7 \times 17.6$ mm$^2$, consisting of $1152 \times 1920$ pixels. We group $144 \times 144$ SLM pixels into a macropixel. The dimension of one macropixel determines the lateral beam size at the fiber proximal facet. The phase of each macropixel is varied independently between 0 and $2\pi$. Each SLM phase pattern comprises $\sim 8 \times 8$ macropixels over an area

covered by the incident Gaussian beam. Reducing the macropixel size would increase the number of macropixels and thus the degree of control of input wavefront, but at the cost of diffraction loss from the abrupt change of phase from one macropixel to the next.

To further enhance the SBS threshold, we optimize the SLM phase pattern. Starting from a random phase modulation, we arbitrarily select a macropixel, scan its phase from 0 to $2\pi$ with a step of $\pi/10$, and evaluate the objective function, e.g., difference between the transmitted signal power and the backward Stokes power, for each phase value. After the scan, the phase of this macropixel is set to the value corresponding to the highest objective function. We continue to optimize another macropixel until all the phases of macropixels are optimized. The SBS threshold is higher after one round of optimization. We then iterate this process by starting another round of optimization. The SBS threshold usually saturates after three iterations with the same objective function, indicating that the optimization has converged to a local maximum of the SBS threshold. We then change the objective function to, e.g., the transmitted signal power, and the threshold enhancement ($P_{\text{th}}/P_{\text{FM}}$) may rise slightly after one to two rounds of optimization.

Finally, we search for the SLM phase pattern for focusing through an MMF, as shown in Fig. 5 of the main text. In order to excite all fiber modes, the axial distance from the focal spot to the distal fiber facet is less than $R\sin\theta$, where $R$ is the fiber core radius, and $\sin\theta$ is the NA. We monitor the intensity distribution on a focal plane with a CMOS camera (Allied Vision Mako GM-419B-NIR) and select the position of focus within the field of view of the MMF. Output intensity at the selected location of focus is used as the objective function for optimizing the input phase pattern. The optimization process begins with a flat phase pattern. We select the macropixel located at the center of the incident Gaussian beam, where the intensity is maximal. We scan the phase of this macropixel from 0 to $2\pi$ and find the value at which the power at the focus is maximal [3]. After setting its phase to the optimal value, we repeat the optimization process for a neighboring macropixel. We progress in a spiral-out trajectory to optimize all macropixels. Due to the smooth variation of the optimized phase over neighboring macropixels, the diffraction loss is small, allowing us to reduce the macropixel size to half of that for random phase modulation. The power at focus typically saturates after one or two rounds of optimization, and the optimized SLM pattern is a smooth phase modulation with $\sim 16 \times 16$ macropixels, as displayed in Fig. S2.

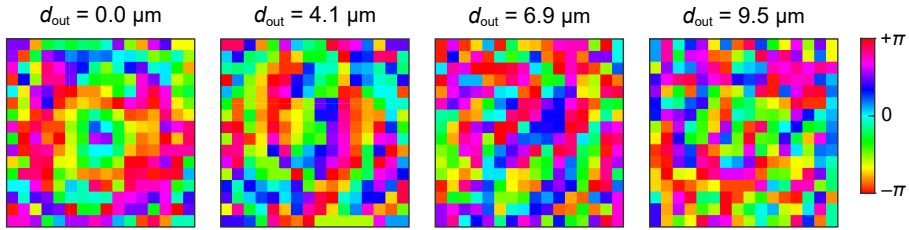

**Fig. S2. Phase modulation for output focusing.** Optimized SLM phase patterns of $16\times16$ macropixels for focusing at different distances to the fiber axis: $d_{\text{out}} = 0.0$, 4.1, 6.9, and 9.5 $\mu$m (left to right), showing smooth phase variations over neighboring macropixels.

4

## 2 Theory and simulation

### 2.1 Multimode excitation by focused input

We simulate input focusing to different positions of the fiber core to calculate the signal mode contents $\{P_l\}$ and predict the corresponding SBS thresholds using our multimode SBS theory. Figure S3a–c shows the numerical estimation of the excited mode contents for (a) FM-only excitation, (b) on-axis focusing and (c) off-axis focusing for multimode excitations, demonstrated experimentally in Fig. 2 of the main text. The effective number of excited modes is defined as $M_{\text{eff}} = (\sum_l P_l)^2 / \sum_l (P_l^2)$. On-axis focusing excites only a few radial HOMs, and $M_{\text{eff}} \approx 4$ [Fig. S3b]. Off-axis focusing excites additional HOMs with non-zero azimuthal index [Fig. S3c], and $M_{\text{eff}}$ increases monotonically with the distance $d_{\text{in}}$ between the focal spot and the fiber axis [Fig. S3d]. Figure S3e shows the predicted SBS threshold enhancement ($P_{\text{th}}/P_{\text{FM}}$) as a function of $M_{\text{eff}}$. From $d_{\text{in}} = 0$ to 8 $\mu$m, $M_{\text{eff}}$ increases gradually from 4 to 30, leading to an increase in the SBS threshold enhancement from 1.4 to 2.3 [Fig. S3e]. Moving beyond $d_{\text{in}} = 8$ $\mu$m, $M_{\text{eff}}$ rapidly rises from 30 to 85 at the fiber core edge ($d_{\text{in}} = 10$ $\mu$m), and the final threshold enhancement of $\sim$3 is predicted [Fig. S3d,e].

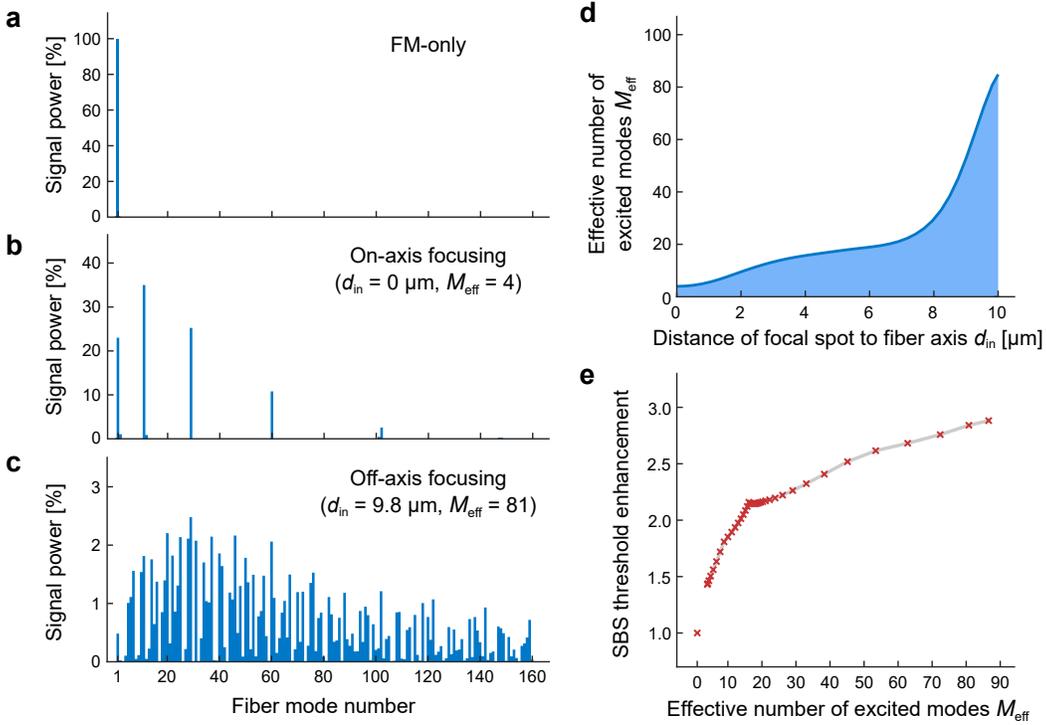

**Fig. S3. Effective number of excited modes vs. SBS threshold. a–c**, Excited signal mode contents at $\lambda = 1064$ nm for FM-only excitation (**a**), on-axis input focusing (**b**), and off-axis input focusing (**c**, near core edge) for a 50-meter-long 20-$\mu$m-core fiber, showing that the number of excited fiber modes increases from condition **a** to **c**. Modes are ordered according to their propagation constants. **d**, Effective number of excited modes ($M_{\text{eff}}$) vs. distance of input focal spot to fiber axis ($d_{\text{in}}$). $M_{\text{eff}} = (\sum_l P_l)^2 / \sum_l (P_l^2)$ increases from 4 to 85 as input focus is moved away from fiber axis. **e**, SBS threshold enhancement $P_{\text{th}}/P_{\text{FM}}$ is raised from 1.4 to 2.8 as $M_{\text{eff}}$ increases from 4 to 85 (*obtained for input focusing).

Due to the exponential growth of Stokes power via SBS, the fiber mode with the highest Brillouin gain will dominate the reflected beam profile above the SBS threshold. The launching condition of the input light to the fiber determines the signal mode content, which in turn selects the Stokes mode with the highest Brillouin gain. In Fig. S4, we use our multimode SBS theory to predict the dominant Stokes modes for various distances of the input focus to the fiber axis, $d_{\text{in}}$, for linearly polarized excitation. When $d_{\text{in}}$ is small (0–3 $\mu$m), mostly radial modes (signal) are excited, providing higher Brillouin gain for radial modes (Stokes). More specifically, the nearly degenerate $HE_{12}$ and $EH_{12}$ modes dominate the backward Stokes in this range [see the left inset in Fig. S4]. For larger $d_{\text{in}}$'s, the signal includes more non-radial modes, leading to higher Brillouin gain for non-radial modes. For $d_{\text{in}} \approx 5$–7 $\mu$m, the lowest-order non-radial modes $HE_{21}$ and $EH_{21}$ begin to dominate the reflected beam profile above

5

the SBS threshold [right inset in Fig. S4]. The theoretical predictions match very closely with the experimentally measured reflection profiles in both cases [Fig. S4]. The far-field intensity pattern of Stokes above the SBS threshold is primarily a coherent superposition of $HE_{12}$ and $EH_{12}$ modes for $d_{in} \approx 0$–$3$ $\mu$m, and of $HE_{21}$ and $EH_{21}$ modes for $d_{in} \approx 5$–$7$ $\mu$m. At $d_{in} \approx 5$ $\mu$m, the dominant modes in Stokes switches, leading to a change of slope in the SBS-threshold-enhancement curve [main panel of Fig. S4].

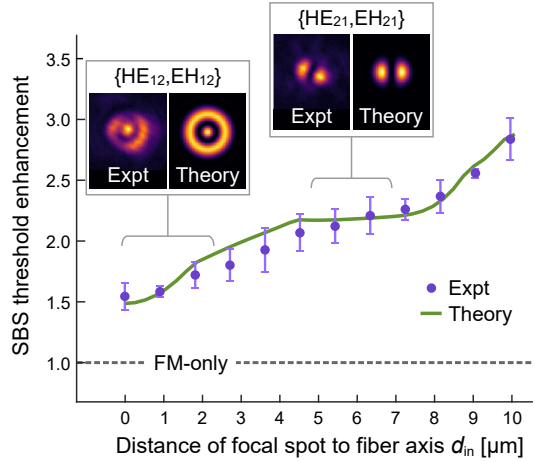

**Fig. S4. Far-field intensity patterns of backward Stokes: theory vs. experiment.** With a linearly polarized focused input, $HE_{12}$ and $EH_{12}$ modes experience the maximum Brillouin gain for the distance of focal spot to fiber axis $d_{in} \approx 0$–$3$ $\mu$m. For $d_{in} \approx 5$–$7$ $\mu$m, the dominant Stokes modes are $HE_{21}$ and $EH_{21}$. These theoretical predictions are consistent with the experimentally measured Stokes profiles for respective $d_{in}$'s. Each profile is a coherent superposition of the HE and EH modes as predicted.

## 2.2 Modeling fiber imperfections

In Fig. 3a of the main text, the SBS threshold enhancement predicted by our multimode SBS theory agrees well with the experimental data. Since the experimental data are influenced by various imperfections in the fiber, such as mode-dependent loss, linear mode coupling, and polarization mixing, we have systematically included these effects in our theoretical model. Below, we provide details for experimental characterization and theoretical modeling of each of these effects individually, along with how they modify the threshold enhancement prediction.

**Mode-dependent loss**
As light propagates through the fiber, some of it will leak out of the core due to imperfect optical confinement. Such loss varies with fiber modes, the higher-order modes suffer more loss due to weaker confinement. To experimentally characterize the mode-dependent loss (MDL) in our fiber, we measure the transmittance as a function of the distance $d_{in}$ of the input focus to the fiber axis. Without MDL, the fiber transmittance is expected to be unity for $d_{in}$ less than the fiber core radius $R$ (dashed blue curve in Fig. S5a), and to drop sharply for $d_{in} > R$. The experimentally measured transmittance (purple crosses) is less than unity for $d_{in} < R$ and continues to drop with increasing $d_{in}$. This is a manifestation of MDL with stronger loss for higher-order modes, because the number and order of excited HOMs increase with $d_{in}$ [Fig. S3b–d]. To model it quantitatively, the loss coefficient of a mode is considered to vary quadratically with its azimuthal index, and the proportionality constant is a fitting parameter [4]. Our phenomenological model can be derived by *ab initio* methods [5, 4, 6] considering the fiber bending/twisting and the scattering due to disorder. The quadratic dependence arises due to loss being proportional to the amount of power striking normal to the core–cladding interface, which depends on the azimuthal angle for the mode. Using this model, we calculate the fiber transmittance and vary the single fitting parameter to minimize the deviation from the measured transmittance for different $d_{in}$. The theoretical fit (solid green curve in Fig. S5a) agrees well with the experimental data, validating our model.

6

Once the MDL in our fiber is characterized experimentally, we include it in our theoretical prediction of the SBS threshold enhancement for different values of $d_{\text{in}}$. In Fig. S5b, the MDL lowers the SBS threshold, because the effective number of excited modes is smaller. When the signal is focused at a larger $d_{\text{in}}$, more higher-order modes are excited, and they suffer stronger loss, leading to a larger drop of the SBS threshold. This result suggests that reducing the MDL can lead to an even higher enhancement of the SBS threshold using multimode excitation.

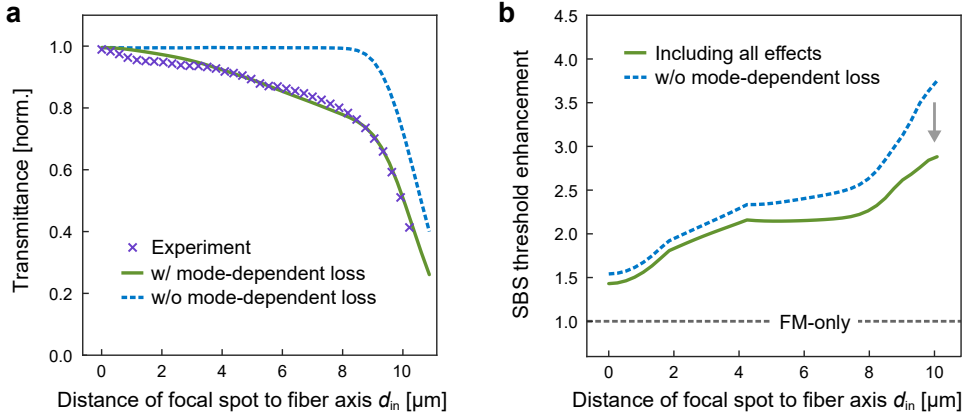

**Fig. S5. Effect of mode-dependent loss on SBS threshold enhancement. a**, Transmittance of light at $\lambda = 1064$ nm through a loosely coiled, 50-meter-long, 20-$\mu$m-core fiber vs. distance of input focal spot to fiber axis $d_{\text{in}}$, showing decrease in transmittance with increasing $d_{\text{in}}$ due to stronger loss for higher order modes. Crosses: experimental data, solid/dashed curve: theoretical curve calculated with/without mode-dependent loss. **b**, Calculated SBS threshold enhancement over FM-only excitation, $P_{\text{th}}/P_{\text{FM}}$, with and without mode-dependent loss as a function of $d_{\text{in}}$. The loss causes a bigger drop of threshold enhancement at larger $d_{\text{in}}$, where more high-order modes are excited and suffer stronger loss. Both linear mode coupling and polarization mixing are included in the theoretical calculation.

**Linear mode coupling**

Without mode coupling, the mode content for a signal remains constant throughout the fiber. However, fiber imperfections and external perturbations cause linear mode coupling. This generically results in an increase in the effective number of excited modes, which raises the SBS threshold. Typically, the linear mode coupling is strongest for the neighboring modes and decreases with the increasing difference in modal propagation constants. We introduce the mode coupling with a banded random matrix shown in Fig. S6a. The matrix element $T_{nl}$ dictates the power in mode $n$ when unity power is input to mode $l$. We construct the coupling matrix $\mathbf{T} = \mathbf{I} + \gamma \mathbf{R}$, where $\mathbf{I}$ is an identity matrix, $\mathbf{R}$ is a banded random matrix with elements between 0 and 1, and $\gamma$ determines the coupling strength. $\mathbf{T}$ is then normalized such that the sum of elements in each row is unity, which ensures power conservation.

Without linear mode coupling, when the input light is focused to the fiber axis ($d_{\text{in}} = 0$), only the FM and some radial HOMs are excited, resulting in a 1.2× enhancement of the SBS threshold. However, experimentally measured far-field intensity pattern of the transmitted light for $d_{\text{in}} = 0$ [Fig. 2b3] reveals a small amount of non-radial HOMs, indicating the presence of weak linear mode coupling in the fiber. The observed SBS threshold enhancement is ∼1.5×, higher than the theoretical prediction without mode coupling. After selecting a banded random matrix $\mathbf{R}$ of bandwidth equal to 5 nondegenerate modes, we fit the value of $\gamma$ such that the SBS threshold enhancement predicted with $\mathbf{T}$ matches the experimental value for $d_{\text{in}} = 0$. The fitting gives $\gamma = 0.03$, confirming weak linear mode coupling in our fiber. For all other values of $d_{\text{in}}$, we use the same coupling matrix $\mathbf{T}$ to correct for the excited mode contents, and then calculate the SBS thresholds.

The results for the SBS threshold enhancement with (solid curve) and without (dashed curve) the linear mode coupling are shown in Fig. S6b. As predicted, linear mode coupling generally increases the SBS threshold, and the increase is maximal for $d_{\text{in}} = 0$ and minimal for the largest $d_{\text{in}}$ ($= 10 \ \mu$m). This is because for a large $d_{\text{in}}$, most modes are already excited at the fiber input, and linear mode coupling causes minimal changes to the mode content throughout the fiber.

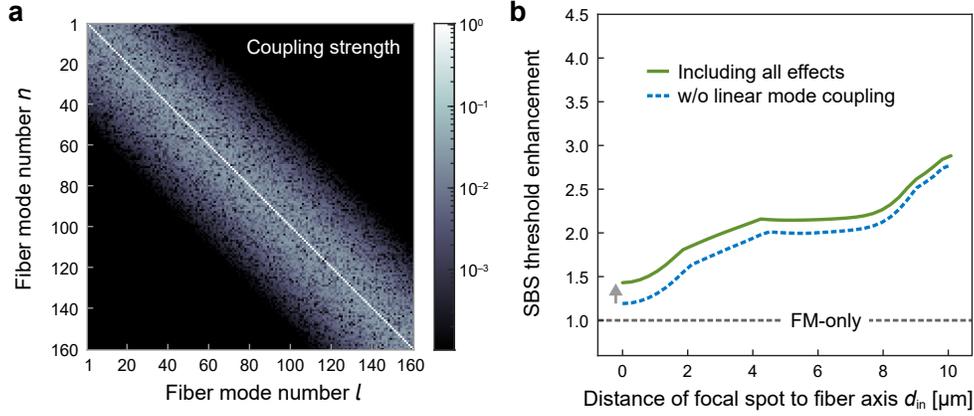

**Fig. S6. Effect of linear mode coupling on SBS threshold enhancement. a**, Banded random matrix **T** for simulating linear mode coupling in the MMF at $\lambda = 1064$ nm, through fitting the experimental data in Fig. 3a of the main text. Coupling strength (plotted in log scale) is larger for modes with smaller difference in their propagation constants. **b**, Calculated SBS threshold enhancement over FM-only excitation, $P_{\text{th}}/P_{\text{FM}}$, with and without linear mode coupling as a function of input focusing distance from fiber axis, $d_{\text{in}}$. Linear mode coupling spreads signal power to more fiber modes and increases the SBS threshold enhancement. The effect is more dramatic at smaller $d_{\text{in}}$, since only a small number of fiber modes ($M_{\text{eff}} < 10$ for $d_{\text{in}} \sim 0$) is excited at the fiber input. Both mode-dependent loss and polarization mixing are included in the theoretical calculation.

## Polarization mixing

Experimentally, linearly polarized light is coupled into a 50-meter-long MMF, and the output polarization state differs from the input one. This is attributed to two separate effects. Consider HOMs with non-zero azimuthal index in the same group, i.e., having identical intensity profiles but different polarization states. Their propagation constants are slightly different. When linearly polarized light excites a superposition of these modes at the fiber input, they will walk off upon propagation in the fiber, producing elliptically polarized light at fiber output. This process effectively causes a power division between two orthogonal polarizations and thus an enhanced SBS threshold for HOMs. We account for this effect in our model by utilizing vector modes instead of linearly polarized modes. In a perfect fiber, such an effect is absent for radially symmetric FM and HOMs. However, fiber imperfections and bending or twisting introduce weak birefringence and polarization mixing, even for the FM. Previous works have shown that, in a single-mode fiber with complete polarization scrambling by imperfection-induced weak birefringence, 1/3 of the input power couples to the FM with the orthogonal polarization to the input [7].

To verify that these two effects are present in our fiber, we use a linear polarizer to characterize the output polarization state. With linearly polarized input, we measure the output intensity while rotating the linear polarizer. The ratio of minimum power $P_{\text{min}}$ to maximum power $P_{\text{max}}$ is plotted against the distance of the input focal spot to the fiber axis, $d_{\text{in}}$. We theoretically predict this dependence by combining the two depolarization effects. Figure S7a shows that the theoretical prediction matches closely with the experimental data. At $d_{\text{in}} = 0$, $P_{\text{min}}/P_{\text{max}} \neq 0$, because the weak birefringence causes depolarization of the FM and a few radial HOMs that are excited. At a larger $d_{\text{in}}$, depolarization is stronger, as a large number of non-radial HOMs are excited.

The polarization mixing reduces the gain for SBS due to power division into two orthogonal polarizations. As a result, the SBS threshold for FM-only excitation is increased. This leads to a lower enhancement of the SBS threshold by multimode excitation, as the enhancement is given by the ratio of the multimode SBS threshold to the FM-only threshold. Figure S7b shows the threshold enhancement with (solid) and without (dashed) polarization mixing. Once the FM-only threshold is corrected for the depolarization effect, a lower enhancement of the SBS threshold reduces the slope of threshold enhancement with $d_{\text{in}}$.

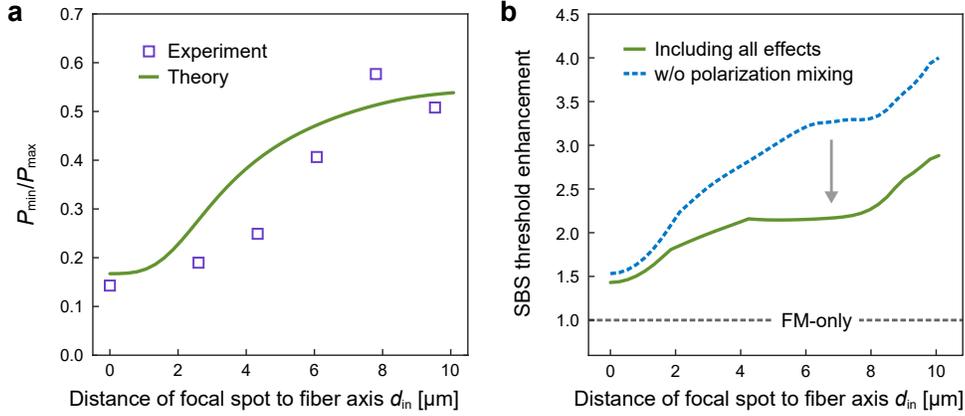

**Fig. S7. Effect of polarization mixing on SBS threshold enhancement. a**, Average ratio of minimum to maximum transmitted power through a linear polarizer at MMF output $P_{\min}/P_{\max}$, when linearly polarized light at $\lambda = 1064$ nm is focused to the fiber input facet at a distance $d_{\mathrm{in}}$ from the core center. Squares: experimental data, solid curve: theoretical estimation. **b**, Calculated SBS threshold enhancement $P_{\mathrm{th}}/P_{\mathrm{FM}}$ with and without polarization mixing as a function of $d_{\mathrm{in}}$. The slope is reduced by polarization mixing, which raises the SBS threshold for FM-only excitation. Both mode-dependent loss and linear mode coupling are included in the theoretical calculation.

## 2.3 Optimization of SBS suppression

**Optimized mode contents**

To gain insight into how an optimized input wavefront enhances the SBS threshold in a multimode fiber, we numerically simulate our wavefront shaping experiment. The same configuration and number of SLM macropixels are set up for phase modulation of a Gaussian beam similar to the CW laser at $\lambda = 1064$ nm. For each SLM phase pattern, the SBS threshold is predicted by our multimode SBS theory including mode-dependent loss, linear mode coupling, and polarization mixing. The predicted threshold is used as the objective function for optimizing the phase modulation. Figure S8a shows an example of the optimization process starting with a random phase pattern. By optimizing the phase of the macropixels one by one, the threshold enhancement increases from 2.6 to 3.5 and then saturates after two to three rounds of optimization of all macropixels. This is in good agreement with the experimental result in Fig. 4b of the main text, and the optimized mode content is shown in Fig. S8b. We repeat the optimization process with different initial phase patterns and present two more examples in Fig. S8b. While the mode contents are different, they all feature a few widely spaced groups of modes. Instead of spreading input power into all modes, the optimization leads to a selective combination of modes that effectively utilizes the inhomogeneous intermodal and intramodal coupling strengths to maximize the SBS threshold.

Figures S8b and 4c reveal that different initial phase patterns and/or different sequences of macropixels for optimization will reach varying mode contents finally, but the corresponding values of SBS threshold enhancement are all around 3.3–3.5 [Fig. 4d]. To quantitatively evaluate how different the optimized mode contents are, we perform 50 optimizations with different initial phase patterns, and calculate the correlation between each pair of optimized mode contents $\{P_l(j)\}$, where $l$ is the mode index, and $j$ denotes the $j^{\mathrm{th}}$ optimization. The Pearson correlation coefficient has a mean value of 0.87 and a standard deviation of 0.07. Thus most optimization processes find similar mode contents from different initial phases. However, the minimum correlation coefficient is merely 0.57. and the mode contents for the least correlated pair are shown in lower panels of Fig. S8b. Nevertheless, the corresponding threshold enhancements are both 3.4, indicating different combinations of excited modes can reach comparable SBS thresholds that are notably higher than the random phase patterns. All optimizations take the same strategy to maximize the SBS threshold: exciting widely spaced groups of modes.

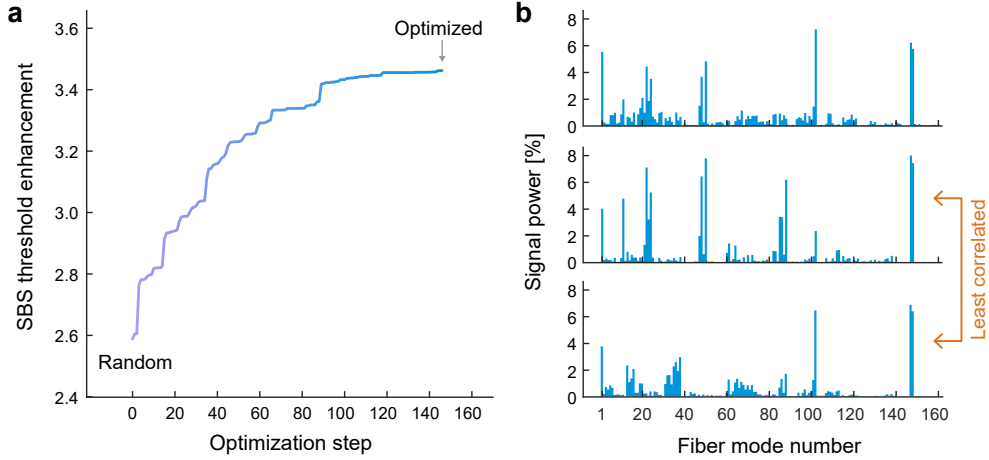

**Fig. S8. Simulated optimization of SBS threshold enhancement. a**, SBS threshold enhancement over FM-only excitation is recorded after optimizing the phase of each macropixel. The optimization starts with a random phase pattern of 8 × 8 macropixels, and the corresponding SBS threshold enhancement is ∼2.6. After two to three rounds of optimization of all macropixels, the threshold enhancement is saturated to ∼3.5. **b**, Three examples of optimized mode contents feature widely spaced groups of modes. Lower two panels are the pair with the least correlated mode contents among 50 optimized ones with different initial random phase patterns and different sequences of macropixel optimization.

### Effective number of excited modes

Figure S9 compares all the approaches employed in this work to realize multimode excitation for SBS suppression. First, tight focusing of input light to the fiber core center excites HOMs and increases the SBS threshold over FM-only excitation. Moving the input focus away from fiber axis further increases the effective number of excited modes $M_{\text{eff}}$ from 4 to 85 (out of 160 modes), leading to a monotonic rise of SBS threshold enhancement up to 2.8. Next, random phase modulation of the input wavefront of linearly polarized light effectively excites 40–60 modes, resulting in SBS threshold enhancement of 2.3–2.7. Finally, wavefront optimization further pushes the threshold enhancement to 3.3–3.5. However, the effective number of modes is lower, $M_{\text{eff}} \sim 30$, illustrating the most efficient way of mitigating SBS is selective mode excitation, not uniform excitation.

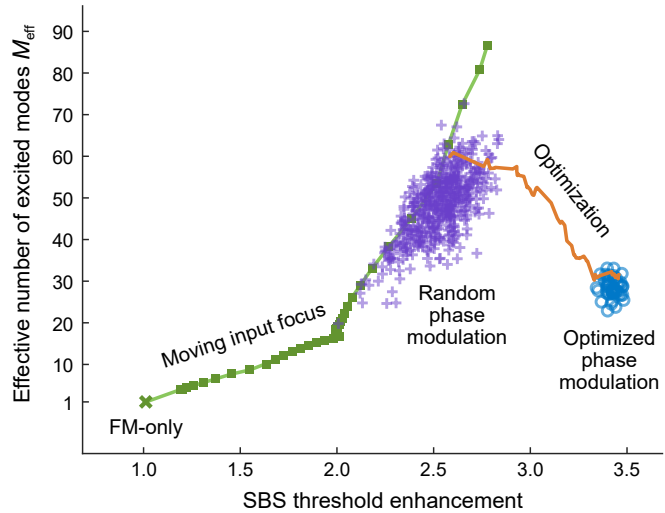

**Fig. S9. SBS threshold enhancement vs. effective number of excited modes.** SBS threshold increases by changing the excitation condition from FM-only (green ×) to multimode excitation with a tightly focused input beam (green ■). Moving the input focus from fiber core center to edge increases threshold up to 2.8. Multimode excitation by random phase modulation (purple +) leads to a SBS threshold enhancement of 3.3–3.5. Both cases display a positive correlation between the SBS threshold enhancement and the effective number of excited modes in the fiber. However, optimization of input phase modulation results in selective modal excitation (blue ○) and a reduction in the effective number of excited modes for further enhancement of the SBS threshold.

**Comparison of SBS suppression schemes**

To maximize the threshold, full control of input field amplitude, phase, and polarization in every fiber mode is needed. In the current experiments, the incident beam on the SLM features a Gaussian profile of its field amplitude. The SLM modulates only the phase front without changing the amplitude profile. Since the SLM corresponds to the far field of the fiber proximal facet, higher intensity near the center of its active region promotes the lower-order modes in the fiber, thus reducing the effective number of excited modes and the SBS threshold. With both amplitude and phase modulation of an input wavefront, we can enhance the HOM contents and the SBS threshold. Our multimode SBS theory predicts a further increase of the SBS threshold enhancement to 4.5 or more upon both amplitude and phase optimization, surpassing the maximum threshold enhancement of 3.5 achieved by phase-only optimization.

## 2.4 Output focusing efficiency

In our experiment, and simulation shown below, the efficiency of focusing light through a multimode fiber is given by the ratio of power within the focal spot to the total transmitted power. The focal spot diameter is equal to twice the full width at half maximum of the intensity. Experimentally, the focusing efficiency is ∼0.7 with phase-only optimization of the input wavefront [Fig. 5 in the main text]. To find the maximum focusing efficiency possible with phase-only modulation of a single polarization, we numerically simulate the output focusing through a multimode fiber [Fig. S10]. We utilize time-reversal symmetry, which is equivalent to phase conjugation for a continuous wave. A diffraction-limited CW source is placed close to the fiber distal facet and propagates through a 50-meter-long fiber to the proximal end. Since many modes are excited, their interference forms a speckle pattern, which is Fourier transformed to the SLM plane. To simulate phase-only modulation by the SLM, we apply a Gaussian profile to the field amplitude. Then we phase-conjugate the field and perform an inverse Fourier transform to obtain the field coupled to the fiber. Finally, we calculate the field transmitted through the fiber and reach the focal plane to compute the focusing efficiency.

Figure S10 (orange curve) shows the (azimuthally averaged) focusing efficiency as a function of the distance of the focal spot to the fiber axis. In our simulation, we ignore mode-dependent loss, linear mode coupling, and polarization mixing in the MMF, and use much smaller macropixels (and thus a smoother phase pattern) than in the experiment. With phase-only modulation of the input wavefront, the focusing efficiency is ∼0.8, and fluctuates slightly with the focus position. This result confirms that the focusing efficiency achieved experimentally is close to the theoretical limit. For comparison, the focusing efficiency with both amplitude and phase modulations is near unity [blue curve in Fig. S10].

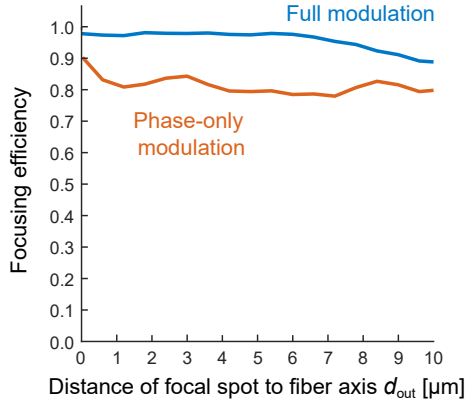

**Fig. S10. Output focusing efficiency by input wavefront shaping.** Compared to both amplitude and phase modulation of input wavefront, the efficiency of focusing light through a multimode fiber decreases from almost 1 to around 0.8 by phase-only modulation. The axial distance of the focal plane from the fiber output facet is 20 μm. Focusing efficiency varies slightly with the distance of the focal spot to the fiber axis, $d_{\text{out}}$, within the field of view.


\end{document}